\documentclass[authoryear]{elsarticle}

\usepackage{graphicx}
\usepackage{epstopdf}
\usepackage{amsfonts}
\usepackage{color}
\usepackage{mathrsfs}
\usepackage{psfrag, amssymb, amsmath, cite}
\usepackage{verbatim}
\usepackage{epsfig} 
\usepackage{times} 
\usepackage{amsmath} 
\usepackage{amssymb}  
\usepackage{amsfonts}
\usepackage[colorlinks,linkcolor=red,anchorcolor=blue,citecolor=blue]{hyperref}
\usepackage{cite}
\usepackage[ruled]{algorithm2e}
\usepackage{fancybox}
\usepackage{framed}
\usepackage{enumitem}

\usepackage{amsthm}
\usepackage{mathtools}

\newtheorem{remark}{Remark}

\newtheorem{theorem}{\textbf{Theorem}}

\newtheorem{lemma}{\textbf{Lemma}}

\renewcommand\footnoterule{\kern-3pt \hrule width 2in \kern 2.6pt}

\journal{Nowhere}

\begin{document}

\begin{frontmatter}



\title{\LARGE A gallery of diagonal stability conditions with structured matrices (and review papers)}


\author{Zhiyong Sun}
\address{Control Systems Group, Department of EE, TU Eindhoven, the Netherlands}


\begin{abstract}
This note  presents a summary and review of various conditions and characterizations for matrix  stability (in particular  diagonal matrix stability) and matrix stabilizability. 
\end{abstract}


\end{frontmatter}


\section{Definitions and notations}
\begin{itemize}
    \item A square real matrix is a \textbf{Z-matrix} if it has nonpositive off-diagonal elements. 
    \item A \textbf{Metzler} matrix is a real matrix in which all the off-diagonal components are nonnegative (equal to or greater than zero). 
    \item A Z-matrix with positive principal minors is an \textbf{M-matrix}.
    \begin{itemize} 
        \item Note: There are numerous equivalent characterizations for M-matrix \citep{fiedler1962matrices, plemmons1977m}.  A more commonly-used condition is the following: A matrix $A \in \mathbb{R}^{n \times n}$ is called an M-matrix, if its non-diagonal entries are non-positive and its  eigenvalues have positive real parts.  
    \end{itemize}
    \item A square matrix $A$ is     \textbf{(positive) stable}  if all its eigenvalues have positive parts. Equivalently, a square matrix $A$ is     (positive) stable iff there exists a positive definite matrix $D$ such that $AD+DA^T$ is positive definite. 
    \begin{itemize}
        \item Note: in control system theory we often define stable matrix as the set of square matrices whose eigenvalues have negative real parts (a.k.a. \textbf{Hurwitz} matrix). The two definitions of stable matrices will be distinguished in the context. 
    \end{itemize}
    \item A square complex matrix is a \textbf{P-matrix} if it has positive principal minors.
    \item A square complex matrix is a \textbf{$P^+_0$-matrix} if it has nonnegative principal minors and at least one principal minor of each order is positive.
    \item A real square matrix $A$ is \textbf{multiplicative D-stable} (in short, \textbf{D-stable}) if $DA$ is stable for every positive diagonal matrix $D$. 
    \item A square  matrix  $A$ is called \textbf{totally stable}   if any principal submatrix of $A$ is D-stable. 
    \item A real square matrix $A$ is said to be \textbf{additive D-stable} if $A + D$ is stable for every nonnegative diagonal matrix $D$.
    \item A real square matrix $A$ is said to be \textbf{Lyapunov diagonally stable}  if there exists a positive diagonal matrix $D$ such that $AD + DA^T$ is positive definite. 
    \begin{itemize}
        \item Note: Lyapunov diagonally stable matrices are often referred to   as just \textbf{diagonally stable} matrices or as \textbf{Volterra–Lyapunov stable},  or as \textbf{Volterra dissipative} in the literature (see e.g., \citep{logofet2005stronger}). 
    \end{itemize}
    \item A matrix $A  = \{a_{ij}\}\in \mathbb{R}^{n \times n}$ is   \textbf{generalized   \textit{\textbf{row}}-diagonally dominant}, if there exists $x = (x_1, x_2, \cdots, x_n) \in \mathbb{R}^n$ with $x_i >0$, $\forall i$, such that
\begin{align}
    |a_{ii}| x_i > \sum_{j=1, j \neq i}^{n} |a_{ij}|x_j, \forall i = 1, 2, \cdots, n.
\end{align}
\item 
A matrix $A  = \{a_{ij}\} \in \mathbb{R}^{n \times n}$ is   \textbf{generalized   \textit{\textbf{column}}-diagonally dominant}, if there exists $x = (x_1, x_2, \cdots, x_n) \in \mathbb{R}^n$ with $x_i >0$, $\forall i$, such that
\begin{align}
    |a_{jj}| x_j > \sum_{i=1, i \neq j}^{n} |a_{ij}|x_i, \forall j = 1, 2, \cdots, n.
\end{align}
\begin{itemize}
    \item Note: the set of generalized   \textit{\textbf{column}}-diagonally dominant matrices is equivalent to the set of generalized   \textit{\textbf{row}}-diagonally dominant matrices \citep{varga1976recurring, sun2021distributed}. They are also often referred to as \textbf{quasi-diagonally dominant} matrices \citep{kaszkurewicz2012matrix}. 
\end{itemize}
    \item For a real matrix $A = \{a_{ij}\} \in \mathbb{R}^{n \times n}$, we associate it with a  \textbf{comparison matrix} $M_A = \{m_{ij}\} \in \mathbb{R}^{n \times n}$, defined by 
    \begin{align}
    m_{ij} =   \left\{
       \begin{array}{cc}
       |a_{ij}|,  &\text{  if  } \,\,\,\,j  = i;  \\ \nonumber
       -|a_{ij}|,  &\text{  if  } \,\,\,\, j  \neq i.   \nonumber  
       \end{array}
      \right.
    \end{align}
    A given matrix $A$ is called an \textbf{H-matrix} if its comparison matrix $M_A$ is an M-matrix. 
    \begin{itemize}
        \item The set of H-matrix is equivalent to the set of quasi-diagonally dominant matrices \citep{kaszkurewicz2012matrix, sun2021distributed}. 
    \end{itemize}
    \item A square matrix $A$ is \textbf{diagonally stabilizable} if there exists a   diagonal matrix $D$ such that $DA$ is stable.
    \end{itemize}
 Note: Many definitions above for real matrices also carry over to complex matrices;  the distinction between real and complex matrices will be made clear in the context. 
    
\section{Conditions for diagonally stabilizable matrices}    

\textit{\textbf{A key motivating question}: Given a square matrix $A$, can we find a diagonal matrix $D$ such that the matrix $DA$ is stable?  }
\\

Fisher and Fuller \citep{fisher1958stabilization} proved the following result:
\begin{theorem} \citep{fisher1958stabilization}
If $P$ is a real $n \times n$ matrix fulfilling the condition:
\begin{itemize} \label{thm:fisher-fuller}
    \item (A): $P$ has at least one sequence of non-zero principal minors $M_k$ of every order
$k = 1,2,\cdots,n$, such that $M_{k-1}$ is one of the $k$ first principal minors of $M_k$;
\end{itemize}
then there exists a real diagonal matrix $D$ such that the characteristic equation
of $DP$ is stable. 
\end{theorem}

The Fisher-Fuller theorem is also formulated as the following alternative version:
\begin{theorem}
Let $P$ be an $n \times n$ real matrix all of
whose leading principal minors are positive. Then there is an $n \times n$ positive
diagonal matrix $D$ such that all the roots of $DP$ are positive and
simple.
\end{theorem}

Fisher later gave a simple proof for a similar yet stronger result \citep{fisher1972simple}.

\begin{theorem} \citep{fisher1972simple}
If $P$ is an $n \times n$ real   matrix that has at least one nested set of principal minors, $M_k$, such that $(-1)^k M_k >0, \forall k = 1\cdots, n$, then there exists a real diagonal
matrix $D$  with positive diagonal elements such that the characteristic roots of
$DP$ are all real, negative, and distinct. 
\end{theorem}
\begin{remark} Some remarks on the conditions of diagonally stabilizable matrices are in order. 
\begin{itemize}
    \item The above theorems involve determining the sign of (at least) one nested set of principal minors. In \citep{johnson1997nested}, sufficient conditions are determined for an $n$-by-$n$ zero-nonzero pattern to allow a nested sequence of nonzero principal minors. In particular, a method is  given to \textit{sign} such a pattern so that it allows a nested sequence of $k$-by-$k$ principal minors with $\text{sign}(-1)^k$ for $k =  1, \cdots, n$. 
    \item The condition in the Fisher-Fuller theorem appears as a \textit{sufficient} condition for matrix diagonal stabilizability. A necessary condition for matrix diagonal stabilizability is: for each order $k = 1\cdots, n$, at least one $k \times k$ principal minor of $P$ is non-zero. It is unclear what would be \textbf{the} necessary and sufficient condition. 
\end{itemize}
\end{remark}

Ballantine \citep{ballantine1970stabilization} extended the above Fisher-Fuller theorem to the complex matrix case. 

\begin{theorem} \citep{ballantine1970stabilization}
Let $A$ be an $n \times n$  \textbf{complex} matrix all of whose leading
principal minors are nonzero. Then there is an $n \times n$ \textbf{complex} diagonal
matrix $D$ such that all the roots of $DA$ are positive and simple.
\end{theorem}
 
\begin{remark}
It is shown in \citep{hershkowitz1992recent} the above Ballantine theorem cannot be strengthened by replacing ``complex
diagonal matrix D'' by ``positive diagonal matrix D''. A counterexample is shown in \citep{hershkowitz1992recent} involving a $2 \times 2$ complex matrix $A$ with positive leading principal minors that there exists no positive diagonal matrix $D$ such that the eigenvalues of $DA$
are positive.
\end{remark}

A related problem to characterize diagonal stabilizable matrix is the \textbf{Inverse Eigenvalue Problem} (IEP), and Friedland \citep{friedland1977inverse} proved the following theorem.
\begin{theorem} \citep{friedland1977inverse}
Let $A$ be a given $n \times n$ \textbf{complex} valued matrix. Assume
that all the principal minors of $A$ are different from zero. Then for any
specified set $\lambda = \{\lambda, \cdots, \lambda_n \} \in \mathbb{C}^n$ there exists a diagonal \textbf{complex} valued
matrix $D$ such that the spectrum of $AD$ is the set $\lambda$. The number of such $D$
is finite and does not exceed $n!$. Moreover, for almost all $\lambda$ the number of the diagonal matrices $D$ is exactly $n!$.
\end{theorem}

\begin{remark}
The Friedland theorem of the IEP problem in the complex matrix case cannot be directly carried over to the real case. Further, it is easy to show with a counterexample of a $2 \times 2$ matrix that eigenvalue positionability
in the real case cannot always be guaranteed, even with nonzero principal minors. 
\end{remark}

In \citep{hershkowitz1992recent} the following two theorems are proved. 
\begin{theorem} \citep{hershkowitz1992recent}
Let $A$ be a \textbf{complex} square matrix with positive leading
principal minors, and let $\epsilon$ be any positive number. Then there exists a positive diagonal matrix $D$ such that the eigenvalues of $DA$ are simple, and
the argument of every such eigenvalue is less in absolute value than $\epsilon$.
\end{theorem}

\begin{theorem} \citep{hershkowitz1992recent}
Let $A$ be a complex square matrix with real principal
minors and positive leading principal minors. Then there exists a positive
diagonal matrix $D$ such that $DA$ has simple positive eigenvalues.
\end{theorem}

\begin{remark}
The above theorems all present certain sufficient conditions to characterize diagonally stabilizable matrix and the IEP problem, and  they are not necessary. A
necessary condition for the diagonal matrix $D$ to exist is that for each order $i$, at least one $i \times i$ principal minor of
A is nonzero. However, a full characterization (with necessary and sufficient condition) for diagonally stabilizable matrix still remains an open problem. 
\end{remark}

A variation of the diagonal matrix stabilization problem is the following:
\begin{itemize}
    \item Problem (*): Given a real square matrix $G$,   find a real diagonal
matrix $D$ such that the product $GD$ is Hurwitz together with all its principal submatrices.
\end{itemize}
Surprisingly, a necessary and sufficient condition exists for solving the above problem as shown in  \citep{locatelli2012necessary}. Let $\mathcal{M}:= \{1, 2, \cdots,  m\}$ and $\mathcal{F}:= \{f | f \subset \mathcal{M}\}$. 
Further, for any $m \times m$ matrix $\Delta$, denote by $\Delta (f)$ the principal submatrix obtained from it after removing the rows and columns with indexes in $f$, $f \in \mathcal{F}$. The main result of \citep{locatelli2012necessary} proves the following:
\begin{theorem} \citep{locatelli2012necessary}
Problem (*) admits a solution if and only if
\begin{align}
    \text{det} (G(f)) \text{det} (G_D(f)) > 0,  \forall f   \in \mathcal{F} 
\end{align}
where $G_D = \text{diag}\{g_{ii}\}$. Moreover, if the above condition  is satisfied, then there exists $\bar \epsilon >0$ such that, for any given $\epsilon \in (0, \bar \epsilon)$, the matrix 
\begin{align}
    D:= G_D Z(\epsilon), Z(\epsilon): =  -\text{diag}\{\epsilon^i \}
\end{align}
solves the stabilization problem (*). 
\end{theorem}

\newpage
\section{Conditions for diagonally stable matrices}  
We give a short summary of available conditions for diagonally stable matrices (excerpts from \citep{barker1978positive}, \citep{cross1978three} and \citep{hershkowitz2006matrix}). 
\begin{itemize}
    \item \citep{barker1978positive} Lyapunov diagonally stable matrices are P-matrices.
    
    \item \citep{barker1978positive} A matrix $A$ being Lyapunov diagonally stable is equivalent to that there exists a positive diagonal matrix $D$ such that $x^TDAx >0$ for   all nonzero vectors $x$. 
    \item \citep{barker1978positive} A $2 \times 2$ real matrix is Lyapunov diagonally stable if and only if it is a P-matrix.
    \item \citep{cross1978three} For a given Lyapunov diagonally stable  matrix $P$,   all principal submatrices of $P$ are Lyapunov diagonally stable.
    \item \citep{barker1978positive} A real square matrix $A$ is Lyapunov diagonally stable if and only if for every nonzero real symmetric positive semidefinite matrix $H$ the matrix $HA$ has at least one positive diagonal element.
    \begin{itemize}
        \item Note: this result is termed the BBP theorem, and is proved again in \citep{shorten2009alternative} with a simpler proof. 
    \end{itemize}
    \item \citep{cross1978three} The set of Lyapunov diagonally stable matrices is a strict subset of multiplicative D-stable matrices.
    \item \citep{cross1978three} The set of Lyapunov diagonally stable matrices is a strict subset of additive D-stable matrices.
        \begin{itemize}
        \item Note:  Multiplicative D-stable and additive D-stable matrices are not necessarily diagonally stable. 
    \end{itemize}
    \item A Z-matrix is Lyapunov diagonally stable if and only if it is a P -matrix (that is, it is an M-matrix). 
    
    \item A non-singular H-matrix with nonnegative diagonal parts is Lyapunov diagonally stable. 
    \item A  quasi-diagonal dominant matrix with  nonnegative diagonal parts is Lyapunov diagonally stable. Note the equivalence of Hurwitz H-matrix and  quasi-diagonal dominant matrix \citep{sun2021distributed}.

\end{itemize}
The following facts are shown in \citep{cross1978three} and \citep{kaszkurewicz2012matrix}:
\begin{itemize}
    \item  For normal matrices and within the set Z, D-stability, additive D-stability, and diagonal stability are all equivalent to
matrix  stability. 
\item  If a matrix $A$ is Hurwitz stable, D-stable, or diagonally stable, then the matrices $A^T$ and $A^{-1}$ also have the corresponding properties.
\end{itemize}

In \citep{shorten2009theorem} Shorten and Narendra showed the following necessary and sufficient condition for matrix diagonal stability (an alternative proof of the theorem of Redheffer via the KYP lemma):
\begin{theorem} \citep{shorten2009theorem} and \citep{redheffer1985volterra}
Let $A \in \mathbb{R}^{n \times n}$ be a Hurwitz matrix with negative diagonal entries. Let $A_{n-1}$ denote the
$[n-1 \times  n-1]$ leading sub-matrix of $A$, and $B_{n-1}$ denote the corresponding block of $A^{-1}$. Then, the
matrix $A$ is diagonally stable, if and only if there is a common diagonal Lyapunov function for the  LTI systems $\Sigma_{A_{n-1}}$ and $\Sigma_{B_{n-1}}$.
\end{theorem}
The above theorem involves finding a  common diagonal Lyapunov function for a set of   LTI systems, which may be restrictive and computationally demanding in practical applications especially when the dimension of the matrix $A$ is large. 

\newpage
\section{Relations of matrix stability and diagonal stability}
The paper \citep{berman1983matrix} characterizes the relations of certain special matrices for matrix diagonal stability. They define
\begin{itemize}
    \item  $\mathscr{A} = \{A: \text{there exists a positive definite diagonal matrix $D$ }\,\text{such that} \,\,AD +DA^T \text{
is positive definite}\}$; \\i.e., $\mathscr{A}$ denotes the set of diagonally stable matrices;
    \item $\mathscr{L} = \{A: \text{there exists a positive definite   matrix  $D$ }\,\,\text{such that} \,\,AD +DA^T \text{
is positive definite}\}$; i.e., $\mathscr{L}$ denotes the set of (positive) stable matrices;
    \item $\mathscr{P} = \{A: \text{the principle minors of $A$} \,\text{
are positive}\}$; \\i.e., $\mathscr{P}$ denotes the set of P-matrices;
    \item $\mathscr{S} = \{A: \text{there exists a positive vector $x$ such that $Ax$ is positive}\}$; \\i.e., $\mathscr{S}$ denotes the set of semipositive
matrices. 
\end{itemize}

\begin{figure} 
\begin{center}
\vspace{-10pt}
\fbox{\includegraphics[width=1\textwidth]{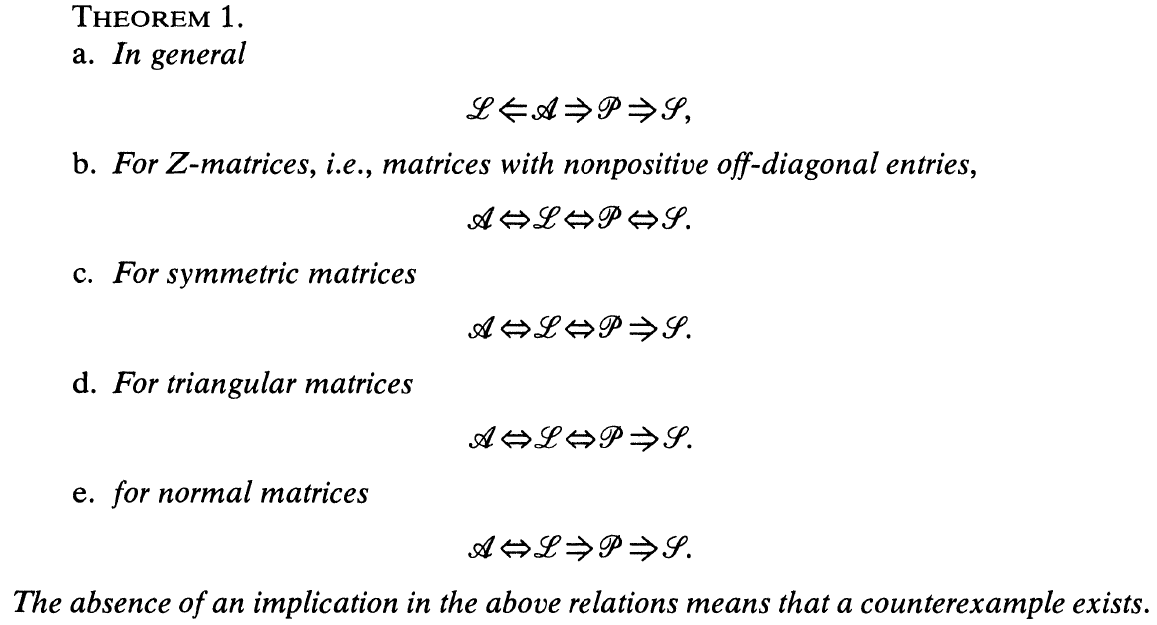}}
\caption{Relations of matrix stability under different matrix types: the main theorem in \citep{berman1983matrix}}
\label{fig:theorem_Ber}
\end{center}
\end{figure}

The main result of \citep{berman1983matrix} is cited and shown in Fig.~\ref{fig:theorem_Ber}. In general, these different sets of structured matrices are not equivalent, and the set $\mathscr{A}$ is a subset of the other sets. However, for Z-matrices, these sets are equivalent. In particular, for the case of Z-matrices, the characterizations of these sets give equivalent conditions for M-matrices (upon a sign change). Note there are yet many more conditions to characterize M-matrices; see e.g., \citep{plemmons1977m}.  

\begin{figure} 
\begin{center}
\vspace{-10pt}
\fbox{\includegraphics[width=0.8\textwidth]{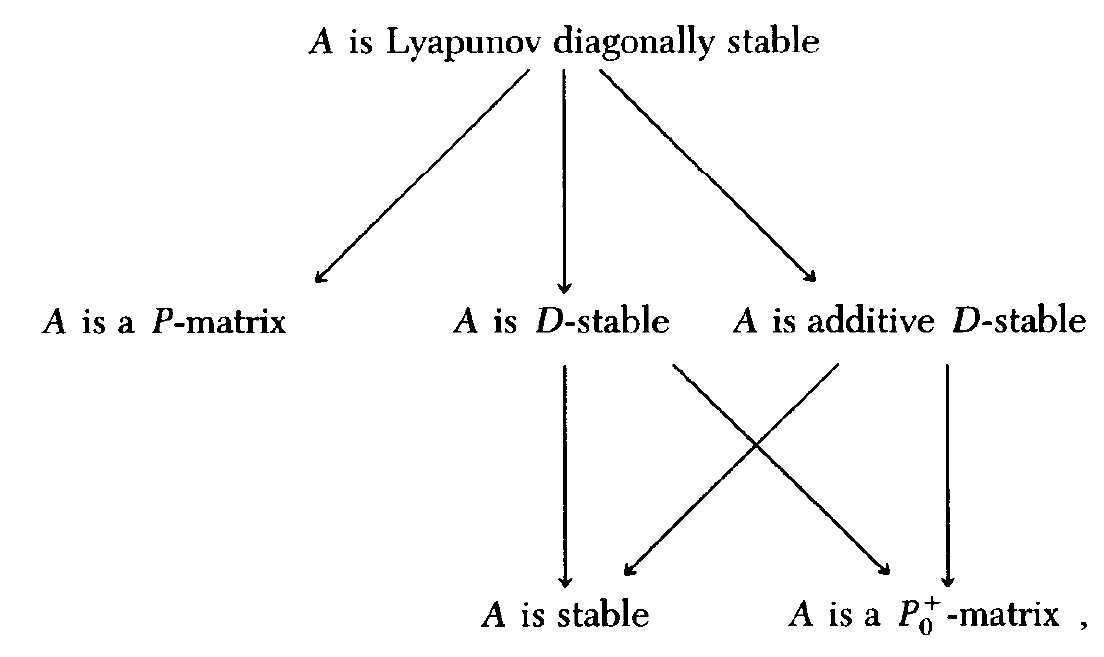}}
\caption{The implication relations between matrix stability conditions, cited from \citep{hershkowitz1992recent}}. 
\label{fig:Relation_Her}
\end{center}
\end{figure} 
\begin{figure}
\begin{center}
\vspace{-10pt}
\fbox{\includegraphics[width=0.8\textwidth]{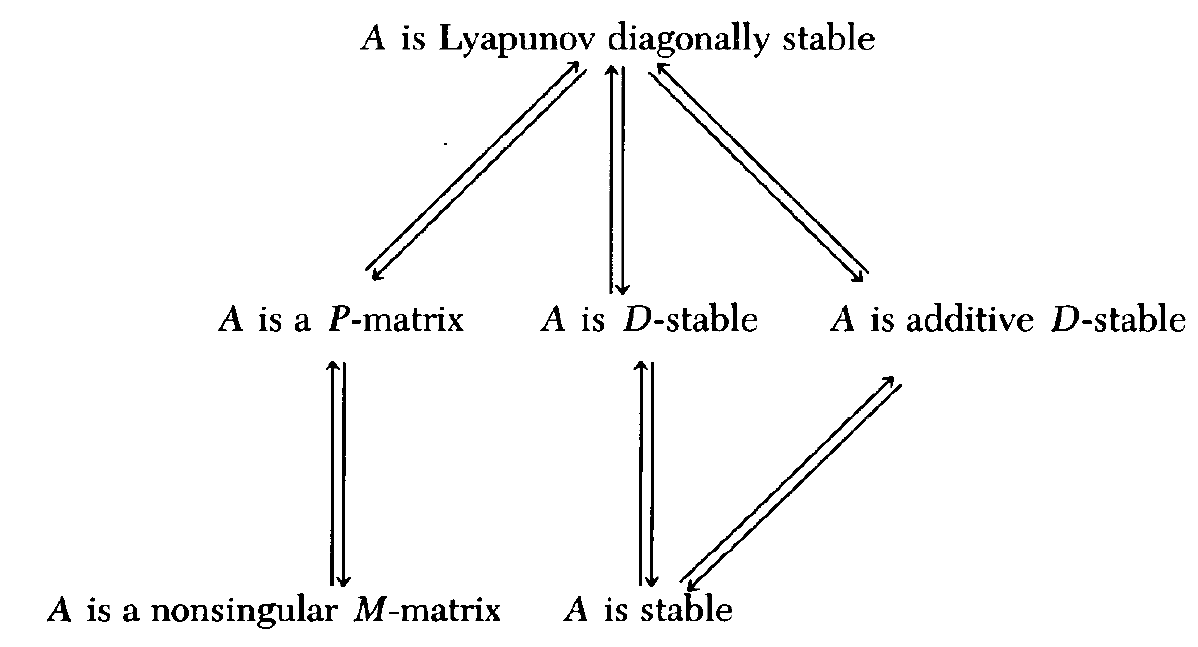}}
\caption{For Z-matrices, all the stability types are equivalent. Cited from \citep{hershkowitz1992recent}}. 
\label{fig:Relation_Her_2}
\end{center}
\end{figure} 

The review paper \citep{hershkowitz1992recent} presents the implication relations between matrix stability conditions, and the equivalent relations of matrix stabilities for Z-matrices, as cited in Figs.~\ref{fig:Relation_Her} and~\ref{fig:Relation_Her_2}. Again, as shown in Figs.~\ref{fig:Relation_Her_2}, for Z-matrices, all the stability types are equivalent. 

\begin{figure} 
\begin{center}
\vspace{-10pt}
\fbox{\includegraphics[width=0.8\textwidth]{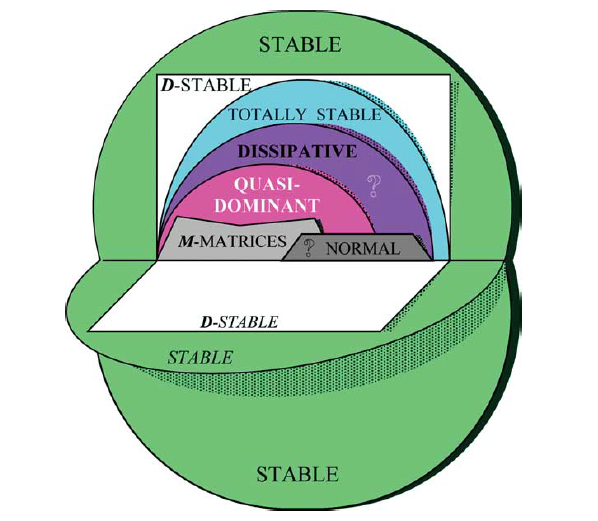}}
\caption{Relations among matrix stabilities. Cited from \citep[Fig.2]{logofet2005stronger}}. 
\label{fig:Relation_Log}
\end{center}
\end{figure}

\begin{figure} 
\begin{center}
\vspace{-10pt}
\fbox{\includegraphics[width=0.8\textwidth]{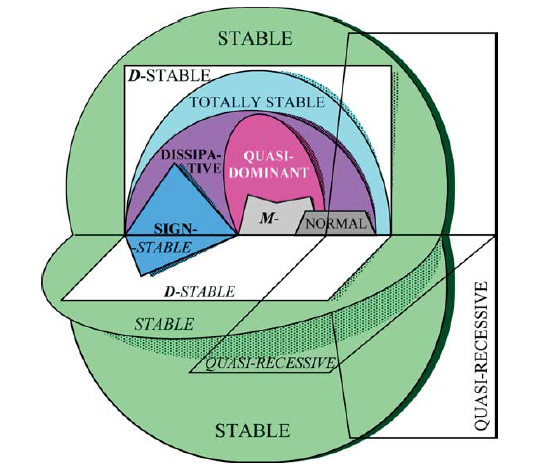}}
\caption{Petals of sign-stable matrices within the Flower. Cited from \citep[Fig.4]{logofet2005stronger}}. 
\label{fig:Relation_Log_2}
\end{center}
\end{figure}

\newpage 
The survey paper \citep{logofet2005stronger} presents some beautiful flower-shaped characterizations of the relations among matrix stabilities, as cited in Figs.~\ref{fig:Relation_Log} and \ref{fig:Relation_Log_2}. 


\newpage
\section{Stability conditions with submatrices and Schur complement}
Stability conditions  of `structured' matrices often involve stability properties of submatrices, which employ block submatrices and their Schur complements to determine stability. 

In  \citep{narendra2010hurwitz}, Narendra and  Shorten presented necessary and sufficient conditions to characterize  if a given Metzler matrix is Hurwitz, based on the fact that a Hurwitz Metzler matrix is  diagonally stable.  These conditions are generalized in \citep{souza2017note}. We recall some main stability criteria from \citep{souza2017note}. 
\begin{lemma} \label{lemma:M-SC}
    Let $A \in \mathbb{R}^{n \times n}$ be a
Metzler matrix partitioned in blocks of compatible dimensions as $A = [A_{11}, A_{12}; A_{21}, A_{22}]$ with $A_{11}$ and $A_{22}$ being square matrices. Then the following statements are equivalent. 
\begin{itemize}
    \item  $A$ is Hurwitz stable.
\item $A_{11}$ and its Schur complement $A/A_{11} :=  A_{22} - A_{21} A_{11}^{-1} A_{12}$ are Hurwitz stable Metzler matrices.
\item $A_{22}$ and its Schur complement $A/A_{22} :=  A_{11} - A_{12} A_{22}^{-1} A_{21}$ are Hurwitz stable Metzler matrices
\end{itemize}
\end{lemma}

\begin{remark} Some remarks are in order. 
\begin{itemize} 
    \item 
For a  structured matrix, the property that its Schur complements also preserve the same stability and structure properties is termed   \textbf{Schur complement stability property}. Other types of structured matrices that have Schur complement stability property include symmetric matrices, triangular matrices, and Schwarz matrices. See \citep{souza2017note}.
    \item  The result on M-matrix in Lemma~\ref{lemma:M-SC} can be generalized to H-matrix: Let $A$ be a H-matrix partitioned in blocks of compatible dimensions as $A = [A_{11}, A_{12}; A_{21}, A_{22}]$ with $A_{11}$ and $A_{22}$ being square matrices. If $A$ is Hurwitz stable, then  $A_{11}$ and its Schur complement $A/A_{11} :=  A_{22} - A_{21} A_{11}^{-1} A_{12}$ are Hurwitz stable H matrices, or $A_{22}$ and its Schur complement $A/A_{22} :=  A_{11} - A_{12} A_{22}^{-1} A_{21}$ are Hurwitz stable H matrices. 

\item Schur complement and its \textbf{closure property} for several structured matrices (including diagonal matrices, triangular matrices, symmetric matrices, P-matrices, diagonal dominant matrices, M-matrices etc.) are discussed in \citep[Chap. 4]{zhang2006schur}. 
\end{itemize}
\end{remark}

\newpage
\section{Application examples of matrix diagonal stability conditions}

The Fisher-Fuller theorem on diagonal matrix stabilizability (Theorem~\ref{thm:fisher-fuller} and its variations) has been rediscovered several times by the control system community, and has been applied in solving distributed stabilization and decentralized control problems in practice. This section reviews two application examples.   
 
\subsection{Conditions for decentralized stabilization}
In \citep{corfmat1973stabilization} Corfmat and Morse solved the following problem:
\begin{itemize}
    \item For given and fixed real matrices $A$ and $P$,   find (if possible) a non-singular diagonal matrix $D$ such that $I+ADP$ is Schur stable (i.e., all eigenvalues of $I+ADP$ are located within the unit circle in the complex plane.
\end{itemize}
To solve the above problem they proved the following:
\begin{theorem}
If $A$ is an $n \times n$ \textbf{strongly non-singular} matrix, then there exists   a
diagonal matrix  $D$ such that $(I + DA)$ is Schur stable. 
\end{theorem}
Note: in \citep{corfmat1973stabilization} a matrix is termed \textbf{strongly non-singular}, if    its all $n$ leading principal
minors are nonzero. 

\begin{theorem}
If $A$ is a fixed non-singular matrix, then there exists a
permutation matrix $P$ such that $PA$ is strongly non-singular. 
\end{theorem}

Solution to decentralized stabilization: the  non-singularity of $A$ is a necessary and sufficient
condition for the existence of a permutation matrix $P$ and a non-singular diagonal matrix $D$ such that $(I + ADP)$ is Schur stable. 

\subsection{Distributed stabilization of persistent formations}
In \citep{yu2009control}, the problem on  persistent formation stabilization involves studying  the stabilizability of the following differential equation
\begin{align*}
    \dot z = \Delta A z
\end{align*}
where $\Delta$ is a diagonal or possibly block diagonal matrix, and $A$ is a rigidity-like matrix on formation shapes. To solve the formation stabilization problem in \citep{yu2009control}
  the following result is employed (\citep[Theorem 3.2]{yu2009control}):
\begin{theorem}
Suppose A is an $m \times m$ non-singular matrix with every leading
principal minor nonzero. Then there exists a diagonal $D$ such that the real parts of
the eigenvalues of $DA$ are all negative.
\end{theorem}
We remark that this is a reformulation of the   Fisher-Fuller theorem. 
\newpage
\section{A selection of key review papers and books on matrix stability and diagonal stability conditions}
\begin{itemize}
    \item  The survey paper \citep{hershkowitz1992recent} that presents a summary of relevant matrix stability results and the developments, up until 1992. 
    \item The paper \citep{bhaya2003characterizations} that presents comprehensive discussions and characterizations for various classes of matrix stability conditions. 
    
    \item The paper \citep{hershkowitz2003positivity} that studies the relations between positivity of principal minors, sign symmetry and stability of matrices. 
 
\item The review paper \citep{hershkowitz2006matrix} that presents an concise overview on matrix stability and inertia. 

\item The book \citep{kaszkurewicz2012matrix} on matrix diagonal stability in systems and computation. 
 
\item The summary paper \citep{logofet2005stronger} that presents a review and some beautiful connections/relations on different matrix stabilities.
 
\item The very long survey paper \citep{kushel2019unifying} that provides a unifying viewpoint on matrix stability, and its historical development. 
 
\item The recent book \citep{johnson2020matrix} on positive matrix, P-matrix and inverse M-matrix. 
 \end{itemize}

\bibliography{adaptive_synchronization}
\bibliographystyle{elsarticle-harv}

\end{document}